\documentclass[conference]{IEEEtran}
\IEEEoverridecommandlockouts
\usepackage{cite}
\usepackage{amsmath,amssymb,amsfonts}
\usepackage{graphicx}
\usepackage{textcomp}
\usepackage{xcolor}

\usepackage{balance}
\usepackage{algpseudocode,algorithm}
\usepackage{float}
\usepackage{flushend}
\usepackage{soul}
\usepackage{cuted} 
\usepackage{siunitx}
\usepackage{booktabs}

\newcommand{\re}{\mathbf{r}_e}          
\newcommand{\rs}{\mathbf{r}^{(s)}}          
\newcommand{\vsk}{\mathbf{v}^{(s)}_k}          
\newcommand{\uhat}{\widehat{\mathbf{u}}}    

\newcommand{\czero}{c}                  
\newcommand{\zb}{\mathbf{z}}
\newcommand{\Zb}{\mathbf{Z}}

\def\BibTeX{{\rm B\kern-.05em{\sc i\kern-.025em b}\kern-.08em
    T\kern-.1667em\lower.7ex\hbox{E}\kern-.125emX}}
\begin{document}

\title{ABC methods for IoT 
Emitter Geolocalisation using LEO Satellite Doppler Measurements\\
}

\author{
\IEEEauthorblockN{Branko Ristic}
\IEEEauthorblockA{\textit{School of Engineering} \\
\textit{RMIT University}\\
Melbourne, Australia\\
branko.ristic@rmit.edu.au}
~\\
\and
\IEEEauthorblockN{Yuna Choi}
\IEEEauthorblockA{\textit{School of Engineering} \\
\textit{RMIT University}\\
Melbourne, Australia  \\
S4199657@student.rmit.edu.au}
~\\
\and
\IEEEauthorblockN{Du Yong Kim}
\IEEEauthorblockA{\textit{School of Engineering} \\
\textit{RMIT University}\\
Melbourne, Australia \\
duyong.kim@rmit.edu.au}
~\\
\and
\IEEEauthorblockN{Akram Hourani}
\IEEEauthorblockA{\textit{Dept. Electr. \& Electron. Eng.} \\
\textit{University of Melbourne}\\
Parkville, Australia \\
akram.hourani@unimelb.edu.au}
}

\maketitle

\begin{abstract}
We address the problem of passive localisation of a stationary, ground-level IoT radio emitter
 using Doppler frequency measurements collected by low-Earth orbit
 (LEO) satellites during an observation window.
The problem is challenging because radio emission from low-cost IoT devices is affected by various compounding sources of measurement
        error, that collectively render the likelihood function intractable in a 
        closed form. Hence, we apply and investigate the performance of  Approximate Bayesian Computation (ABC) methods for this task. Numerical results demonstrate the statistical and computational performance of two ABC methods, rejection sampling ABC and sequential Monte Carlo ABC.
\end{abstract}

\begin{IEEEkeywords}
Approximate Bayesian Computation;
Emitter geolocalisation;
Likelihood-free inference.
\end{IEEEkeywords}

\section{Introduction}

The passive geolocalisation of uncooperative radio emitters is a problem of
significant practical interest in electronic intelligence, border protection,
search-and-rescue, and spectrum management ~\cite{scales_84,hendy2026beyond}. 
A particularly relevant scenario
involves the use of low-Earth orbit (LEO) satellites as passive observers: as
a satellite passes overhead, it receives a Doppler-shifted version of the
emitter's signal, and the resulting time series of frequency measurements
carries information about the emitter's position on the surface of the Earth \cite{nguyen2016algebraic, ellis2020use}.

When the emitter is a precision-engineered device with a well-characterised
oscillator and when the propagation environment is benign, the Doppler
localisation problem can be approached with analytical methods~\cite{nguyen2016algebraic,chestnut_82, anderson_2020, ellis2020use,farzan_2020,hashim2022satellite}. The
zero-crossing time of the S-shaped Doppler curve constrains the emitter to a
cone about the satellite's velocity vector, and the shape of the curve
provides additional geometric constraints. With multiple satellites, the
intersection of these constraints yields a position estimate.

The focus of this paper is on a particularly challenging class of emitters: low-cost IoT
devices. Passive localisation could be required for one of the following reasons:   the IoT emitter is a legitimate telemetry transmitting IoT device, but its registered location is unknown or wrong; 
the IoT device is transmitting without authorisation and there is a requirement to localise it without alerting; 
the IoT could be a distress beacon transmitting from an unknown location (search-and-rescue scenario); 
the IoT device is causing interference on a frequency it should not be using (a spectrum management scenario).

IoT devices are typically equipped
with uncompensated crystal oscillators (XO), and hence are characterised by: a large and unknown initial frequency offset; 
    a time-varying frequency drift driven by temperature fluctuations
          and oscillator aging; 
    no cooperative frequency reference, making the true carrier
          frequency  unknown to any external observer.
These characteristics, combined with ionospheric and tropospheric propagation
delays and satellite ephemeris uncertainty, make the marginal likelihood
function of the observed Doppler measurements, given the emitter position, intractable.
This likelihood function cannot be written in closed form
because it requires marginalisation over several high-dimensional and
partially unknown nuisance processes. 
For this reason we apply {\em likelihood-free}  inference ~\cite{cranmer2020frontier}, in which the likelihood is never evaluated explicitly but instead
approximated by running a stochastic forward simulator many times. In particular, we consider two Approximate Bayesian Computation (ABC) methods \cite{marin2012approximate}: the sample rejection method ABC and the sequential Monte Carlo ABC \cite{toni2008approximate,DelMoral2012}.

\section{Problem formulation}
IoT emitter is modelled as a stationary point source on the surface of the
Earth, with position vector $\re =[\phi_e,\; \lambda_e,\;h_e]^{\!\top}$ parameterised by geodetic latitude $\phi_e$, longitude
$\lambda_e$, and altitude $h_e$ above the WGS84 reference ellipsoid.
The
corresponding Earth-Centred Earth-Fixed (ECEF) Cartesian position vector
 is obtained via the standard WGS84 coordinate
transformation.
The emitter is designed to transmit a continuous-wave signal on a nominal carrier frequency $f_0$, which is assumed known to the observers. Owing to oscillator imperfections, however, the actual instantaneous transmitted frequency deviates from $f_0$ by a small unknown time-varying amount, described in Sec. \ref{s:noise_sources}.

Each satellite $s=1,\dots,S$ is modelled as a circular Keplerian orbit at altitude $h^{(s)}$, characterised by inclination $i$, right ascension of the ascending node (RAAN) $\Omega$, and
initial true anomaly $\nu_0$. The true satellite position $\rs_k$ at discrete-time $t_k$ is obtained by propagating the
orbital elements through the perifocal frame, rotating to the
Earth-Centred Inertial (ECI) frame using the standard rotation matrix
~\cite{curtis2020}, and then transforming to ECEF by accounting
for Earth rotation.

\subsection{Measurement model}
\label{s:meas_model}

The ideal (noise-free) Doppler shift observed by satellite $s$ at $t_k$ is ~\cite{misra2006}
\begin{equation}
    D^{(s)}_k = -\frac{f_0}{\czero}\,
                   \vsk \cdot \uhat^{(s)}_k,
    \label{eq:doppler_geom}
\end{equation}
where $\czero$ is the speed of light,
$\vsk$ is the ECEF satellite velocity vector, and
\begin{equation}
    \uhat^{(s)}_k = \frac{\rs_k - \re}
                      {\|\rs_k - \re\|}
\end{equation}
is the unit line-of-sight vector from emitter to satellite $s$. The scalar
$\vsk \cdot \uhat^{(s)}_k$, featuring in (\ref{eq:doppler_geom}), is the range rate.
The resulting Doppler
curve has the characteristic S-shape
over the duration of a pass, with the zero-crossing
occurring at closest approach.

A satellite is considered to produce a valid measurement only when its
elevation angle as seen from the emitter exceeds a minimum threshold of
$5^\circ$~\cite{misra2006}. 

\subsection{Measurement noise}
\label{s:noise_sources}

Observed Doppler at sensor $s$ is corrupted by noise:
\begin{equation}
    z_k^{(s)} =  D^{(s)}_k + \eta_k^{(s)},
\end{equation}
where $\eta_k^{(s)}$ is a superposition of four independent sources:
\begin{equation}
    \eta_k^{(s)} =   \delta f_k
                      + b^{(s)}_k
                      + e^{(s)}_k
                      + n^{(s)}_k,
    \label{eq:obs_model}
\end{equation}
The noise term $\delta f_k$ is the clock drift (common to all satellites). The IoT emitter is equipped with an uncompensated crystal oscillator (XO)
with no temperature compensation~\cite{vig1991aging, riley2008} and instantaneous
frequency deviation from the nominal $f_0$ modelled as: 
    $\delta f_k = \delta f_0 + w_k$.
The term $\delta f_0$ is a large, unknown initial frequency offset reflecting the XO's absolute frequency accuracy:
\begin{align}
    \delta f_0 & \sim \mathcal{U}(-\Delta f_{\max},\, +\Delta f_{\max}).
    \label{eq:df0}
\end{align}
 The term
$w_k$ is a zero-mean Wiener process ~\cite{gardiner2009}.
Both $\delta f_0$ and the time-varying $w_k$ are
unknown to the localisation algorithm. They constitute a high-dimensional,
correlated nuisance that cannot be marginalised analytically.

The noise term $b^{(s)}_k$ in (\ref{eq:obs_model}) is the atmospheric bias. As the satellite signal travels from the emitter through the atmosphere,
ionospheric and tropospheric refraction introduce a range-equivalent delay
that varies with the elevation angle  and with
time-varying atmospheric conditions.
 The atmospheric base processes at  satellites are unknown to the estimator.

The noise term $e^{(s)}_k$ in (\ref{eq:obs_model}) is the ephemeris-induced Doppler error due to satellite ephemeris uncertainty. The satellite position used by the estimator is derived from Two-Line Element
(TLE) sets propagated with the SGP4 model~\cite{vallado2006revisiting,kelso2007}. A TLE file is only updated
periodically, and SGP4 ignores many small perturbing forces, so the
predicted satellite position diverges from the truth over time. The ephemeris-induced Doppler error is  a structured, geometry-dependent contribution that peaks near
closest approach.

Finally, $n^{(s)}_k$ represents additional measurement uncertainty caused by thermal noise at the receiver. It is modelled as
 zero-mean white Gaussian
additive noise with standard deviation $\sigma_n$ on each Doppler measurement.

\subsection{Testing scenario}
\label{sec:ts}

The testing scenario is summarised in Table~\ref{tab:scenario}. The
emitter is placed on the ground in central Australia. 
Three LEO satellites observe the emitter during a \SI{15}{\minute} window, 
resulting in $K = 900$ time steps at $\Delta t = \SI{1}{\second}$.

\begin{table}[htb]
    \centering
    \caption{Simulation scenario parameters.}
    \label{tab:scenario}
    \begin{tabular}{ll}
        \toprule
        Parameter & Value \\
        \midrule
        Emitter latitude          & $-22.7^\circ$  \\
        Emitter longitude         & $133.9^\circ$ \\
        Emitter altitude          & \SI{30}{\metre} (ground level) \\
        Carrier frequency $f_0$   & \SI{437.525}{\mega\hertz} \\
        Number of satellites      & 3 \\
        Orbital altitude $h^{(s)}; s=1,2,3$         & \SI{550}{\kilo\metre} \\
        Simulation duration       & \SI{900}{\second} (\SI{15}{\minute}) \\
        Sampling interval $\Delta t$      & \SI{1}{\second} \\
        Minimum elevation         & $5^\circ$ \\
        Max. frequency offset $\Delta f_{\max}$     &   \SI{8750}{\hertz}  \\
        Thermal noise std $\sigma_n$ & \SI{3}{\hertz} \\
        \bottomrule
    \end{tabular}
\end{table}
Orbital parameters for the three satellites are listed in
Table~\ref{tab:orbits}. All orbits are circular at $h_{\mathrm{s}} = \SI{550}{\kilo\metre}$. The parameters were selected to ensure
each satellite passes over the emitter within the 15-minute
observation window, at different times and from different approach directions.

\begin{table}[htb]
    \centering
    \caption{Orbital parameters for the three LEO satellites.}
    \label{tab:orbits}
    \begin{tabular}{lccc}
        \toprule
        & Inclination & RAAN & init. true anom.\\
        Satellite & $i$ [deg] & $\Omega$ [deg] & $\nu_0$ [deg]  \\
        \midrule
        SAT-1 & 97.6 & 320.1 & 175.7 \\
        SAT-2 & 98.0 & 328.7 & 185.1 \\
        SAT-3 & 97.8 & 130.5 & -55.3  \\
        \bottomrule
    \end{tabular}
\end{table}

\begin{figure}[htb]
    \centering
    \includegraphics[trim=50pt 50pt 50pt 50pt, clip, width=0.48\textwidth]{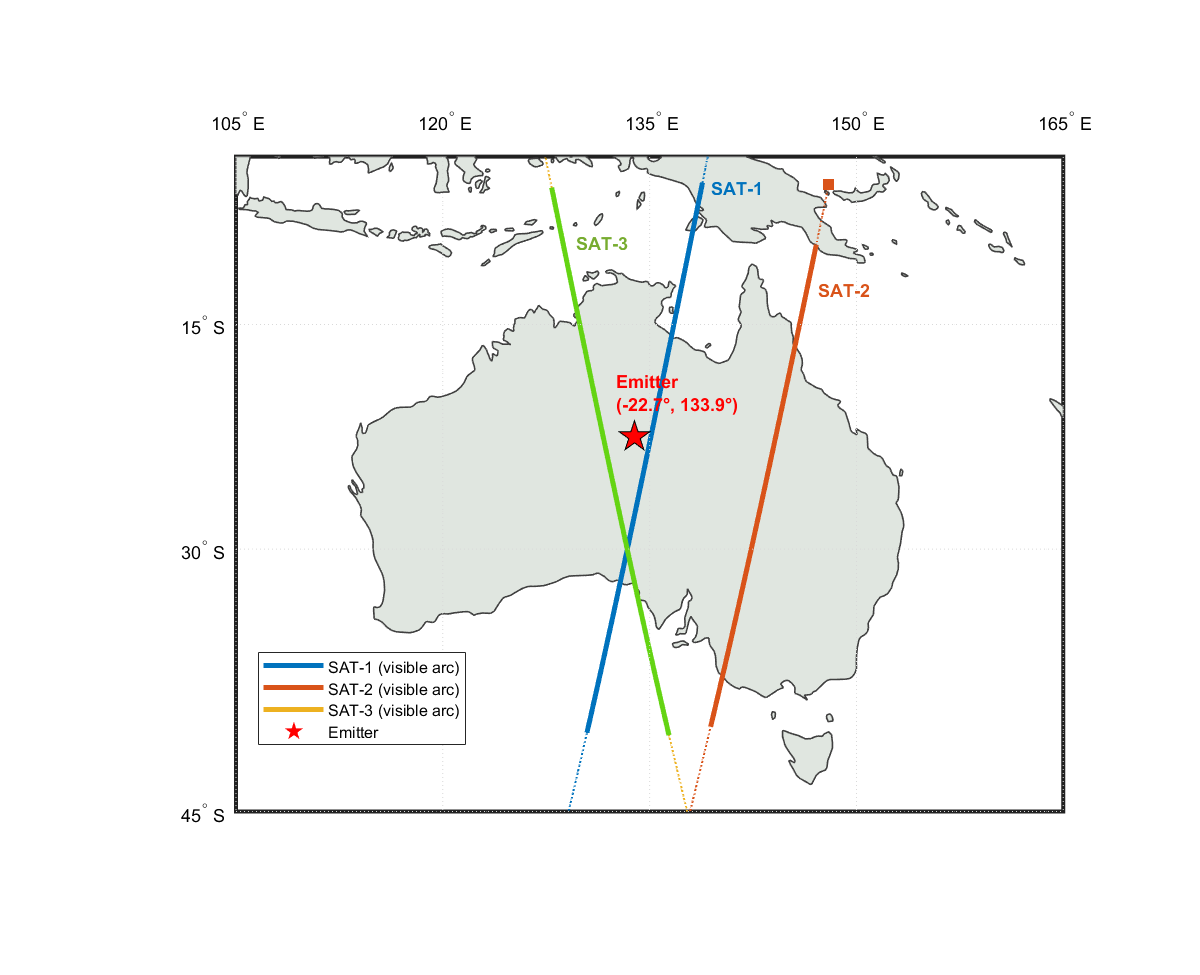}
    \caption{Satellite ground tracks and emitter location.}
    \label{fig:map}
\end{figure}

\begin{figure}[htb]
    \centering
    \includegraphics[width=0.49\textwidth]{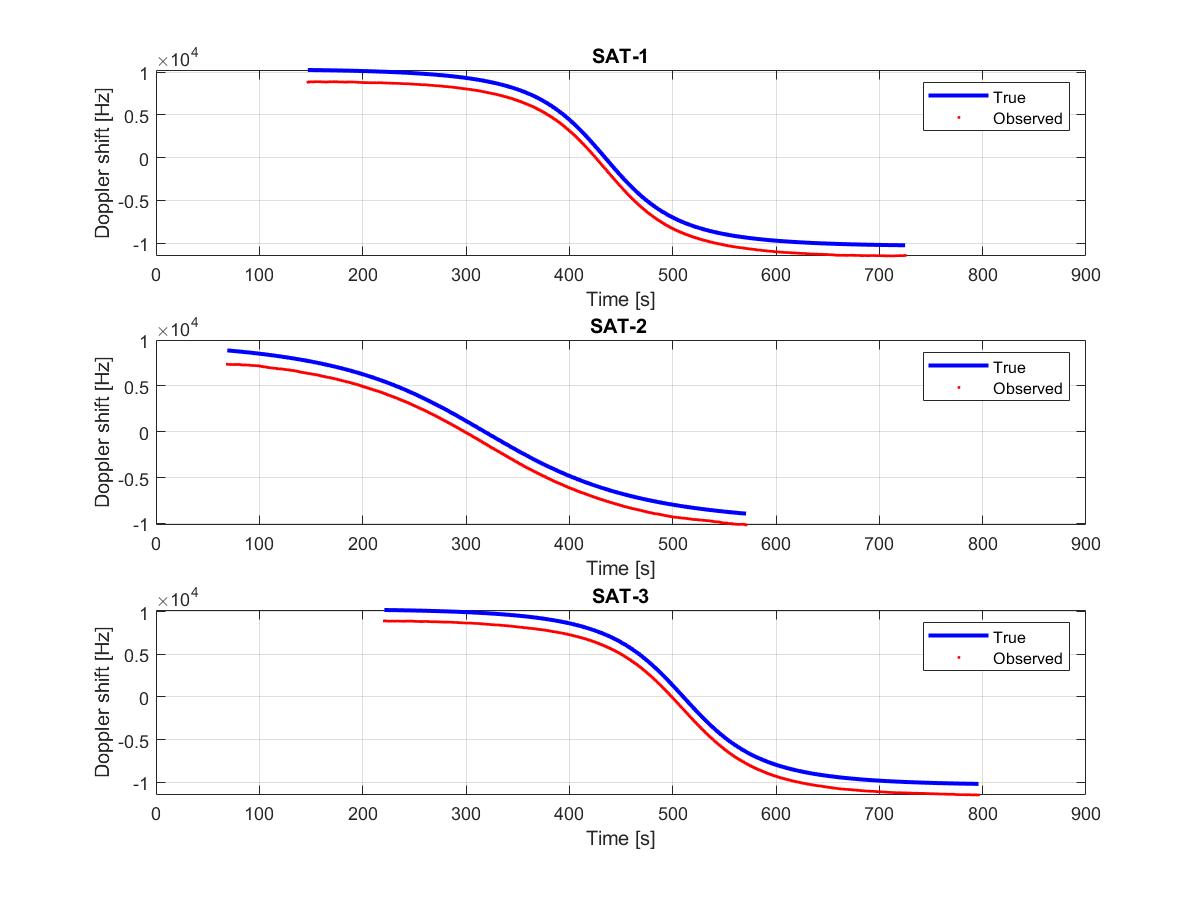}
    \caption{Ideal (blue) and observed (red) Doppler
             time series for all three satellites.}
    \label{fig:dopplers}
\end{figure}

Figure~\ref{fig:map} shows the ground tracks of the three satellites over the
\SI{15}{\minute} window. Solid lines indicate the visible arc (elevation
$> 5^\circ$). Figure~\ref{fig:dopplers} shows the ideal (noise-free) and observed Doppler time series for all three satellites. The  Doppler time series
follow the characteristic S-shape of a LEO pass over the
visible arc, with the transition rate determined by the satellite's angular
velocity relative to the emitter. The three passes have different transition
timings and rates, reflecting the different geometries of Table~\ref{tab:orbits}. The visible offset in Doppler between the two curves is dominated by the unknown initial frequency offset $\delta f_0$ of
the IoT crystal oscillator. 

\subsection{Problem statement}

The objective is to infer the emitter position $\re$ from the observed Doppler time series
$z^{(s)}_{1:K}\equiv z_1^{(s)},\dots,z^{(s)}_K$, $s=1,\dots,S$.   A classical Bayesian approach would compute
the posterior~\cite{kay1993statistical,gelman1995bayesian}
\begin{equation}
    p(\re \mid z^{(1)}_{1:K},\dots,z^{(S)}_{1:K}) \propto p(  z^{(1)}_{1:K},\dots,z^{(S)}_{1:K}\mid \re)\, p(\re),
\end{equation}
where $p(\re)$ is a geographic prior. Assuming  conditional independence of Doppler observations on different satellite $p( z^{(1)}_{1:K},\dots,z^{(S)}_{1:K}\mid \re) = \prod_{s=1}^S p( z^{(s)}_{1:K}\mid \re)$ is the
joint likelihood function. The problem is simplified by prior knowledge that the emitter is on the ground. 

The likelihood functions $p( z^{(s)}_{1:K}\mid \re)$, for $s=1,\dots,S$, could be approximated by marginalising the joint
distribution over all nuisance variables (various noise source parameters and inaccurate satellite positions). However, this would be cumbersome as it would require high-dimensional Monte Carlo integration over the joint space of all nuisance
variables. 

\section{Localisation via Likelihood-free inference}

Likelihood-free inference (LFI) methods are used when a mathematical likelihood function is too complex, computationally intractable, or entirely unknown to evaluate analytically. Instead of evaluating the likelihood, LFI relies on a forward generative simulator, such as the one described in Secs. \ref{s:meas_model}, \ref{s:noise_sources}. 
Given any candidate emitter position $\re^*$, one
can draw independent realisations of all nuisance variables from their
respective priors, propagate them through the measurement model, and produce a synthetic Doppler time series
$z_{1:K}^{(s)}(\re^*)$, $s=1,\dots,S$. 
If the synthetic data closely matches the observed experimental data, candidate $\re^*$ is considered plausible.

In this paper we focus on a class of LFI methods collectively referred to as
approximate Bayesian computation (ABC) methods.  In particular, we investigate ABC rejection sampling  and  sequential Monte Carlo ABC algorithms. 

\subsection{Distance function}
\label{sec:abc_distance}

In order to measure how closely synthetic data matches the observed experimental data, we need to introduce the notion of distance function. 

A naive choice of distance --- the Euclidean norm between the observed
and simulated Doppler time series --- fails for this problem, because the observed and
simulated measurements each contain an independent realisation of the
oscillator frequency offset, see (\ref{eq:df0}) and Fig. \ref{fig:dopplers}.  The raw residual between the two
series is therefore dominated by the difference of the two offsets ---
a random variable with a range of $2\Delta f_{\max} = 17.5$~kHz that
carries no position information --- rather than by the geometric
mismatch attributable to the candidate position. 

Note first that visible time steps of the observed
and simulated Doppler time series are in general different. 
Let us introduce the jointly visible time
steps for satellite $s$ as $\mathcal{V}^{(s)} = \mathcal{V}^{(s)}_{\mathrm{obs}} \cap
\mathcal{V}^{(s)}_{\mathrm{sim}}$. Furthermore, let us denote the measurement vectors corresponding to joint visible time steps $\mathcal{V}^{(s)}$ of observed and simulated Doppler as $\zb^{(s)}_{\mathrm{obs}}$ and $\zb^{(s)}_{\mathrm{sim}}$, respectively.

The remedy follows from the structure of the measurement model: over a
single observation window, $\delta\!f_0$ enters as a pure additive
constant. Consequently, all vertical translations
$\{\zb^{(s)}_{\mathrm{sim}} + c : c \in \mathbb{R}\}$ of a simulated data
represent the same position hypothesis. The appropriate distance
between the observed time series and the hypothesis is the distance to the
\emph{closest member} of this equivalence class,
\begin{equation}
  d^{(s)} = \min_{c \,\in\, \mathbb{R}}
  \bigl\lVert \zb^{(s)}_{\mathrm{obs}} -
  (\zb^{(s)}_{\mathrm{sim}} + c\,\mathbf{1}) \bigr\rVert ,
  \label{eq:abc_distance}
\end{equation}
where  $\mathbf{1}$ is the
all-ones vector. The minimisation in \eqref{eq:abc_distance} is
ordinary least squares in the single parameter $c$: writing
$\mathbf{d} = \zb^{(s)}_{\mathrm{obs}} - \zb^{(s)}_{\mathrm{sim}}$, the
objective $J(c) = \sum_{k \in \mathcal{V}^{(s)}} (d_k - c)^2$ is a convex
parabola whose minimiser is the sample mean,
\begin{equation}
  \widehat{c}^{(s)} \;=\; \frac{1}{|\mathcal{V}^{(s)}|}
  \sum_{k \in \mathcal{V}^{(s)}} \bigl( \zb^{(s)}_{\mathrm{obs}}[k] -
  \zb^{(s)}_{\mathrm{sim}}[k] \bigr).
  \label{eq:abc_offset}
\end{equation}
The aligned distance
is exactly invariant to both observed and simulated Doppler offset realisations.
The slowly varying part of the clock drift, the Wiener
process $w_k$ (see Sec. \ref{s:noise_sources}), is not constant and is therefore not removed by the
alignment; its wander about the window mean remains in the residual and acts as correlated noise.

The per-satellite contribution is the root-mean-square of the aligned
residual,
\begin{equation}
  D^{(s)} = \frac{1}{\sqrt{|\mathcal{V}^{(s)}|}}
  \bigl\lVert \zb^{(s)}_{\mathbf{obs}} - \zb^{(s)}_{\mathrm{sim}} -
  \hat{c}^{(s)}\mathbf{1} \bigr\rVert ,
  \label{eq:abc_rmse}
\end{equation}
where the normalisation by $|\mathcal{V}^{(s)}|$ makes contributions
from passes of different durations commensurable. Invoking the
conditional independence of the measurements across satellites, the
overall distance is a superposition:
\begin{equation}
  \mathcal{D} = \sum_{s \,:|\mathcal{V}^{(s)}|\, \ge N_{\min}} D^{(s)} ,
  \label{eq:abc_dist_sum}
\end{equation}
subject to two guards that address failure modes introduced by the
alignment itself. First, satellites with fewer than $N_{\min}$ jointly
visible steps are excluded
from \eqref{eq:abc_dist_sum}: a short arc near the horizon is nearly
flat, so after vertical alignment its residual is deceptively small
for \emph{any} candidate position, and its inclusion would dilute the
distance with uninformative terms. Second, the distance is computed using   \eqref{eq:abc_dist_sum} only
if at least $S_{\min}$ satellites survive this threshold; otherwise $D$ is set to  $\infty$. 


\subsection{ABC rejection sampling}
\label{sec:abc}

For simplicity, let us introduce notation $\Zb_{\mathbf{obs}} \equiv \{z^{(s)}_{1:K}$, $s=1,\dots,S\}$ for observed data. 

Approximate Bayesian computation (ABC) 
exploits the forward simulator described in Secs. \ref{s:meas_model} and \ref{s:noise_sources} to generate  synthetic (simulated) Doppler data for a candidate emitter position $\re$, denoted $\Zb_{\mathbf{sim}}(\re) \equiv 
\{z_{1:K}^{(s)}(\re), s=1,\dots,S\}$.

ABC rejection sampling algorithm  draws the samples from the
approximate posterior:
\begin{equation}
  p_\epsilon(\mathbf{r}_e \mid \Zb_{\mathbf{obs}} ) \;\propto\;
  p(\mathbf{r}_e) \,
  \Pr\bigl\{ \mathcal{D}(\Zb_{\mathbf{obs}}, \Zb_{\mathrm{sim}}(\re)) \le \epsilon
  \;\big|\; \mathbf{r}_e \bigr\},
  \label{e:approx}
\end{equation}
where $p(\mathbf{r}_e)$ is prior distribution. Approximate posterior (\ref{e:approx})
converges to the true posterior as $\epsilon \to 0$ when the
distance is computed on sufficient statistics. The entire inferential
burden therefore shifts to the design of the distance function, which
for this problem was explained in
Sec.~\ref{sec:abc_distance}.

The prior $p(\mathbf{r}_e)$ is uniform over a geographic bounding box
covering Australia, $\phi_e \in [-44^\circ, -10^\circ]$,
$\lambda_e \in [112^\circ, 154^\circ]$, with the altitude known and fixed at
ground level ($h_e = 30$~m). Pseudo-code is given in
Algorithm~\ref{alg:abc}. Note that in line 5, the simulator draws all nuisance parameters internally, and that output $\mathbf{V}_{\mathrm{sim}} = = \{\mathcal{V}^{(s)}_{\mathrm{sim}}; s = 1,\dots,S\}$ contains  visibility masks of simulated Doppler time series.

\begin{algorithm}[t]
\caption{ABC rejection sampler}
\label{alg:abc}
\begin{algorithmic}[1]
\State \textbf{Input:} 
 \Statex Observed Doppler $\Zb_{\mathbf{obs}}$ 
 \Statex Satellite trajectories $\mathbf{R}$
 \Statex Visibility
  masks $\mathbf{V}_{\mathrm{obs}} = \{\mathcal{V}^{(s)}_{\mathrm{obs}}; s = 1,\dots,S\}$;
  \Statex Tolerance
  $\epsilon$;
  \Statex Sample count $N$
\State $n \gets 0$
\While{$n < N$}
  \State Draw $\mathbf{r}_e^{*} \sim
    \mathcal{U}(\text{bounding box})$ \Comment{propose from prior}
  \State $\{\Zb_{\mathbf{sim}}, \mathbf{V}_{\mathrm{sim}}\}
    \gets \textsc{Simulate}(\mathbf{r}_e^{*}, \mathbf{R})$
  \State Compute $\mathcal{D}$ from
    \eqref{eq:abc_distance}--\eqref{eq:abc_dist_sum}
  \If{$\mathcal{D} \le \epsilon$}
    \State $n \gets n+1$; store $\mathbf{r}_e^{(n)} =
      \mathbf{r}_e^{*}$
  \EndIf
\EndWhile
\State \textbf{Output:} Posterior samples
  $\{\mathbf{r}_e^{(n)}\}_{n=1}^{N}$
\end{algorithmic}
\end{algorithm}

The output of Algorithm~\ref{alg:abc} are accepted samples $\{\mathbf{r}_e^{(n)}\}_{n=1}^{N}$, which constitute an
equally weighted particle approximation of
$p_\epsilon(\mathbf{r}_e \mid \Zb_{\mathbf{obs}})$. The IoT emitter position estimate is
taken as the posterior mean, with the sample standard deviations in
latitude and longitude providing a dispersion measure. 


The main drawback of ABC rejection sampler is that the acceptance rate is very
low, since a proposal is accepted only when the candidate position is
consistent with the observed curve shapes under \emph{all} nuisance
realisations drawn in that simulation. This inefficiency of rejection
ABC with a broad prior motivates the sequential Monte Carlo variant
(SMC-ABC), in which a population of candidates is propagated through a
decreasing tolerance schedule.


\subsection{SMC-ABC}

The rejection sampler of Section~\ref{sec:abc} proposes every candidate
from the prior, so its acceptance rate is fixed by the ratio of the
posterior to the prior volume. Sequential Monte Carlo ABC (SMC-ABC)
\cite{toni2008approximate, DelMoral2012} removes this inefficiency by propagating
a population of $N$ candidates through a decreasing sequence of
tolerances $\epsilon_1 > \epsilon_2 > \dots > \epsilon_T$, so that
proposals at each stage are drawn from the neighbourhood of the
surviving population rather than from the full prior.

The initial generation (generation 1) is drawn from
the prior with infinite tolerance. This is carried out by the application of Algorithm 1 with generation 1 tolerance $\epsilon_1 = \infty$, meaning that a candidate need only pass the visibility guards of
Section~\ref{sec:abc_distance}. At each subsequent generation $t=2,3,\dots$, the
tolerance is set adaptively to the $\alpha$-quantile  of the previous generation's distances \cite{DelMoral2012}. New candidates  are generated by sampling a
parent from the weighted population and perturbing it with a Gaussian
kernel whose covariance is twice the weighted population covariance
\cite{Beaumont2009}; proposals falling outside the prior support are
rejected before any simulator call. An emitter position candidate $\mathbf{r}^{(i)}_{e,t}$ is accepted if its
distance does not exceed $\epsilon_t$, and receives the standard
population Monte Carlo importance weight
\begin{equation}
  w^{(i)}_t \;\propto\;
  \frac{p(\mathbf{r}^{(i)}_{e,t})}
  {\sum\limits_{j=1}^{N} w^{(j)}_{t-1}\,
   K\!\bigl(\mathbf{r}^{(i)}_{e,t} \mid
   \mathbf{r}^{(j)}_{e,t-1}\bigr)},
  \label{eq:pmc_weight}
\end{equation}
where the perturbation kernel $K$ is the Gaussian density
$K(\mathbf{r}_e \mid \mathbf{r}'_e) =
\mathcal{N}(\mathbf{r}_e;\, \mathbf{r}'_e,\, \boldsymbol{\Sigma}_t)$,
with covariance $\boldsymbol{\Sigma}_t$ equal to twice the weighted
covariance of generation $t-1$ \cite{Beaumont2009}, i.e. the same
kernel used to generate the proposals.

The recursion terminates when a
common target tolerance $\epsilon_{\mathrm{T}}$ is reached.
Pseudo-code is given in
Algorithm~\ref{alg:smcabc}.

\begin{algorithm}[tbh]
\caption{SMC-ABC}
\label{alg:smcabc}
\begin{algorithmic}[1]
\State \textbf{Input:} 
 \Statex Observed Doppler $\Zb_{\mathbf{obs}}$ 
 \Statex Satellite trajectories $\mathbf{R}$
 \Statex Visibility
  masks $\mathbf{V}_{\mathrm{obs}} = \{\mathcal{V}^{(s)}_{\mathrm{obs}}; s = 1,\dots,S\}$;
  \Statex Target tolerance
  $\epsilon_{\mathrm{T}}$;
  \Statex Sample count $N$
  \Statex Quantile $\alpha$
  \State {\raggedright Obtain Generation $t=1$:\par} {\raggedright $\bullet$ $\{\mathbf{r}^{(i)}_{e,1} \}_{i=1}^N$ by calling Algorithm 1 with $\epsilon_1=\infty$;\par}
 {\raggedright $\bullet$ Save from Algorithm 1 distances $\mathcal{D}^{(i)}_1$; $i=1\dots,N$ \par} 
{\raggedright $\bullet$ Set importance weights: $w^{(i)}_1 = 1/N$; $i=1\dots,N$ \par}
\For{$t = 2, 3, \dots$}
  \State $\epsilon_t \gets \max\bigl(
    \mathrm{quantile}_\alpha(\{\mathcal{D}^{(i)}_{t-1}\}),\;
    \epsilon_{\mathrm{T}} \bigr)$
  \State $\boldsymbol{\Sigma}_t \gets 2 \times$ weighted covariance of
    $\{\mathbf{r}^{(i)}_{e,t-1}, w^{(i)}_{t-1}\}$
    \For{$i = 1, \dots, N$}
    \Repeat
      \State Sample parent $j$ with probability $w^{(j)}_{t-1}$
      \State $\mathbf{r}^{*}_{e} \gets
        \mathbf{r}^{(j)}_{e,t-1} + \boldsymbol{\xi}$,\; where 
        $\boldsymbol{\xi} \sim \mathcal{N}(\mathbf{0},
        \boldsymbol{\Sigma}_t)$ 
      \State $\{\Zb_{\mathbf{sim}}, \mathbf{V}_{\mathrm{sim}}\}
    \gets \textsc{Simulate}(\mathbf{r}_e^{*}, \mathbf{R})$
\State Compute distance
        $\mathcal{D}^{*}$;\; 
    \Until{$\mathcal{D}^{*} \le \epsilon_t$}
    \State $\mathbf{r}^{(i)}_{e,t} \gets \mathbf{r}^{*}_{e}$;
      \; $\mathcal{D}^{(i)}_t \gets \mathcal{D}^{*}$;\;
      $w^{(i)}_t \gets$ Eq.~\eqref{eq:pmc_weight}
  \EndFor
  \State Normalise $\{w^{(i)}_t\}$
  \State \bf{If} $\epsilon_t \le \epsilon_{\mathrm{T}}$, 
    \textbf{break}; \bf{end }
\EndFor
\State \textbf{Output:} weighted population
  $\{\mathbf{r}^{(i)}_{e,t}, w^{(i)}_t\}_{i=1}^{N}$
\end{algorithmic}
\end{algorithm}

The output of Algorithm~\ref{alg:smcabc} are weighted samples $\{\mathbf{r}^{(i)}_{e,t}, w^{(i)}_t\}_{i=1}^{N}$,  approximating the posterior
$p_\epsilon(\mathbf{r}_e \mid \Zb_{\mathbf{obs}})$ at target tolerance $\epsilon = \epsilon_{\mathrm{T}}$. The emitter position estimate is
the weighted sample mean. A dispersion measure is the weighted sample standard deviations in
latitude and longitude. 

\section{Numerical results}

Simulations were conducted using the scenario described in Sec. \ref{sec:ts}.
The following parameters were adopted:  for distance function $N_{\min} = 100$, $S_{\min} = 2$. For Algorithm 1 (rejection ABC) the sample count is $N=100$ and tolerance  is set to $\epsilon = 400$ Hz. For Algorithm 2 (SMC-ABC) we set $N = 256$,  $\epsilon_{\mathrm{T}} = 150$ Hz and  $\alpha = 0.5$. Different parameters we used for Algorithm 1 and Algorithm 2 because rejection sampling ABC is significantly slower to run.

Monte Carlo runs were conducted with the true emitter position fixed (Table \ref{tab:scenario}). All nuisance parameters are redrawn per run, at random.

Considering that the height of the emitter is known, in error analysis we only focus on emitter geodetic latitude and longitude.  Because degrees of latitude and longitude
are incommensurable units, the positional error vector will be expressed in  the local tangent plane, with units in kilometres, east and north of the true position. Let the error vector of the $m$th Monte Carlo run be defined then as 
$\mathbf{e}_m =  [\,(\lambda_e - \widehat{\lambda}_{e,m})\, L_\lambda
\cos\widehat{\phi}_{e,m},\; (\phi_e - \widehat{\phi}_{e,m})\, L_\phi\,]^{\!\top}$,
where $\widehat{\phi}_{e,m}$ and $\widehat{\lambda}_{e,m}$ are the posterior mean latitude  and longitude obtained from the $m$th Monte Carlo run, $L_\phi \approx 111.0$~km/deg and $L_\lambda \approx 111.32$~km/deg
\cite{Farrell1999TheGP}. The root-mean-square error (RMS) in emitter position (in kilometres) is then computed from $M$ runs as 
\[ e_{\mathrm{\text{\tiny RMS}}} = \sqrt{\frac{1}{M}\sum_{m=1}^M \left\lVert \mathbf{e}_m \right\rVert^2}.   \]

Point estimation accuracy alone does not validate a Bayesian estimator: the
\emph{reported uncertainty} must also match the actual errors. An estimator which satisfies this property is referred to as being {\em efficient} \cite{BarShalom2001}. A suitable metric for testing the efficiency by Monte Carlo runs is the normalised estimation error squared (NEES) \cite{BarShalom2001}, defined for the $m$th runs as:
\begin{equation}
  \mathcal{E}_m \;=\;
  \mathbf{e}_m^{\!\top}
  \mathbf{P}_m^{-1}
  \mathbf{e}_m,
  \label{eq:nees}
\end{equation}
where $\mathbf{P}_m$ is the posterior sample covariance of the $m$th run,  in the same east/north local tangent  frame in kilometres (i.e. using $\cos\widehat{\phi}_{e,m}$ and $L_\phi$ and $L_\lambda$). Under the assumption that the estimation error is Gaussian and $\mathbf{P}_m$ correctly reflects the error covariance, NEES $\mathcal{E}_m$ is chi-square distributed with 2 degrees of freedom (denoted $\chi^2_2$).  Across $M$ runs, we report the mean
NEES $ \bar{\mathcal{E}} = \frac{1}{M} \sum_{m=1}^M\mathcal{E}_m$, and we test it against 
$\chi^2_{2M}/M$ acceptance interval chosen based on a given significance level. 

The cost-efficiency of the two algorithms needs to measure  how many simulator runs the sampler has to spend, on average, to generate one statistically independent draw from the posterior. The raw count of {\em simulator calls per run}  makes the rejection ABC look better than it really is. A more honest measure of cost-efficiency is the count of {\em calls per effective size}, defined as the ratio between the {\em simulator calls per run} and the {\em effective sample size} (ESS) $\nu = \left[\sum_i w^{(i)}_t\right]^2 / \sum_i [w^{(i)}_t]^2$.

\begin{table*}[tbh]
  \centering
  \caption{Comparison of the two ABC samplers obtained from 30 Monte Carlo runs (each): ABC rejection
  ($\epsilon = 400$~Hz, $N=100$) and SMC-ABC
  ($\epsilon_{\mathrm{T}} = 150$~Hz, $N = 256$). }
  \label{tab:abc_comparison}
  \begin{tabular}{lcc}
    \toprule
     & Rejection ABC & SMC-ABC \\
    \midrule
    Tolerance $\epsilon$ [Hz]            & 400 & 150 \\
    RMS error [km]        & 6.68 & 4.49 \\
    Mean NEES ($95\%$ acceptance interval)         & 0.104 (1.35--2.78) & 0.553 (1.35--2.78) \\
    Simulator calls per run              & $\sim$363\,000 & $\sim$77\,500 \\
    Calls per effective sample           & $\sim$3\,600 & $\sim$310 \\
    \bottomrule
  \end{tabular}
\end{table*}



Table~\ref{tab:abc_comparison} summarises the performance of both ABC samplers, obtained from $M=30$ Monte Carlo run (of each algorithm).
 Note first  that SMC-ABC dominates the ABC rejection
sampler on every measure: (a) It is more accurate, with RMS error of 4.5~km versus 6.7~km; (b) Its mean NEES is closer to the acceptance interval at significance level of $95\%$ (1.35-2.78): 0.55  versus 0.11;  (c) It is an order of
magnitude cheaper in terms of the count of calls per effective sample. The second observation is that \emph{neither} sampler
is overconfident (in 60 Monte Carlo runs, no single run produced a
NEES above the 95\% acceptance limit). Finally,  note that both estimated posteriors are conservative:
the mean NEES falls significantly below the efficiency interval in both ABC
algorithms, meaning that the reported uncertainty is {\em higher} than the actual geolocation error. 
On this last note,  the conservatism of the reported uncertainty is an
inherent property of likelihood-free inference (rather than a tuning artifact) for this measurement
model at this nuisance level. This conservatism is a benign failure for intended applications, with the error
exclusively in the safe direction. 

\section{Conclusions and future work}


The paper  formulated the passive geolocalisation of a stationary, ground-level IoT emitter from LEO-satellite Doppler measurements as a likelihood-free inference problem, and solved it with two approximate Bayesian computation samplers: rejection ABC and SMC-ABC. The intractability of the likelihood arises directly from the low-cost nature of the emitter — an uncompensated crystal oscillator with a large unknown frequency offset and time-varying drift — compounded by atmospheric propagation delays and satellite ephemeris uncertainty. The central methodological ingredient enabling ABC in this setting is an offset-invariant distance function that projects out the unknown oscillator offset by aligning observed and simulated Doppler series over their jointly visible arcs, so that the residual reflects geometric mismatch attributable to the candidate position rather than the nuisance offset. Across Monte Carlo trials with the true position fixed and all nuisance variables redrawn per run, SMC-ABC outperforms  the rejection sampler on every measure considered. 

Two directions are planned for future work. First, a systematic sensitivity analysis with respect to satellite availability and observation geometry. Localisation accuracy and posterior calibration are expected to depend strongly on the number of visible satellites, the diversity of their approach directions, and the elevation profiles of their passes. Characterising this dependence would clarify the operational envelope of the method and inform observation scheduling.

Second, the incorporation of machine-learning methods for simulation-based inference. 
Neural simulation-based inference — for instance, learning a surrogate for the posterior, the likelihood, or the likelihood ratio directly from simulator output, or learning informative summary statistics in place of the ABC-designed distance function — offers a route to amortised inference that reuses simulation effort across queries and may substantially reduce the number of simulator calls while also reducing the uncertainty in the posterior.

\bibliographystyle{IEEEtran}
\bibliography{references}
\end{document}